\documentclass[letterpaper, 10 pt, conference]{ieeeconf}
\IEEEoverridecommandlockouts 
\usepackage{makeidx}  
\usepackage{graphicx}
\usepackage{amsmath}
\usepackage{amssymb}
\usepackage{xparse}
\usepackage{xspace}
\usepackage{xcolor}
\usepackage{subcaption}
\usepackage{cite}
\graphicspath{{pics/}}
\usepackage{float}
\usepackage{algorithm2e}

\newfloat{Model}{tbp}{ext}

\NewDocumentCommand{\ipic}{mO{scale=0.4}}{
	\includegraphics[#2]{#1.pdf}
}

\NewDocumentCommand{\pic}{O{}mO{scale=0.4}}{
	\begin{figure}
		\center
		\ipic{#2}[#3]
		\caption{\label{#2}	#1}
	\end{figure}
}

\NewDocumentCommand{\pico}{O{}mO{scale=0.4}}{
	\begin{figure*}
		\center
		\ipic{#2}[#3]
		\caption{\label{#2}	#1}
	\end{figure*}
}

\NewDocumentCommand{\pics}{omo}{
	\begin{figure}
		\center
		#2
		\IfValueTF{#1}{\caption{#1 
			\IfValueTF{#3}{\label{#3}}{}
		}}{}		
	\end{figure}
}

\NewDocumentCommand{\subpic}{O{scale=0.4}mm}{
	\subcaptionbox{#2 
		\label{#3}		
	}{\ipic{#3}[#1]}
}

\NewDocumentCommand{\sym}{smO{}moo}{
	\expandafter\NewDocumentCommand\csname #2\endcsname{#3}{\ensuremath{{#4}}\xspace}
}
\NewDocumentCommand{\name}{mmm}{
	\expandafter\NewDocumentCommand\csname #1n\endcsname{}{#2\xspace}
	\expandafter\NewDocumentCommand\csname #1a\endcsname{}{#3\xspace}	
	\expandafter\NewDocumentCommand\csname #1ns\endcsname{}{#2s\xspace}
	\expandafter\NewDocumentCommand\csname #1as\endcsname{}{#3s\xspace}	
}
\sym{func}[mmm]{#1: #2 \rightarrow #3}
\sym{powerset}[mO{}]{\mathcal{P}_{#2}(#1)}
\sym{reals}{\mathbb{R}}[real numbers]
\sym{preals}{\reals^{+}}[positive real numbers]
\sym{intervals}[O{\reals}]{\mathcal{I}(#1)}
\sym{intervalsp}[O{\preals}]{\intervals[\preals]}[set of all intervalls where the lower bound is positive]
\sym{interval}[O{x}O{y}]{[#1, #2]}[an interval where $x$ and $y$ are mostly in \reals]
\sym{naturalnumbers}{\mathbb{N}}[natural numbers]

\sym{buses}[O{}O{}]{N_{#1}^{#2}}[$= \{1, \ldots n\}$, set of buses]
\sym{numofbuses}{n}[number of buses, \cardi{\buses}]

\sym{gen}[O{}O{}]{G_{#1}^{#2}}
\sym{generators}[O{}]{\buses_{\gen}^{#1}}

\sym{load}[O{}O{}]{L_{#1}^{#2}}
\sym{loads}[O{}]{\buses_{\load}^{#1}}

\sym{pa}[O{}O{}]{\Theta_{#1}^{#2}}
\sym{padiff}[O{}O{}]{\Delta_{#1}^{#2}}

\sym{lines}[O{}O{}]{E_{#1}^{#2}}
\sym{linesw}[O{}]{\lines_{#1}'}

\sym{sus}[O{}O{}]{B_{#1}^{#2}}
\sym{suscap}[O{}]{B_{#1}^{*}}

\sym{edge}[O{ab}O{}O{}]{{#1}_{#3}^{#2}}

\sym{flow}[O{}O{}]{F_{#1}^{#2}}
\sym{capa}[O{}O{}]{C_{#1}^{#2}}

\sym{net}[O{}O{}]{\mathcal{N}_{#1}^{#2}}[network]

\sym{mpf}[O{\net}O{}]{\mathcal{MPF}_{#2}(#1)}
\name{mpf}{maximum potential flow}{MPF}

\sym{mf}[O{\net}O{}]{\mathcal{MF}_{#2}(#1)}
\name{mf}{maximum flow}{MF}

\sym{mff}[O{\net}O{}]{\mathcal{MFF}_{#2}(#1)}
\name{mff}{maximum FACTS flow}{MFF}

\sym{mvf}[O{\net}O{}]{\mathcal{MVF}_{#2}(#1)}
\name{mvf}{maximum variable flow}{MVF}

\sym{sols}{\mathcal{T}_{\net}}
\sym{sol}{(\sus, \pa, \flow, \gen, \load)}
\name{sol}{feasible solution}{MPF}

\sym{gr}[O{}O{}]{\mathit{GR}_{#1}^{#2}}
\sym{bg}[O{}O{}]{\mathit{BG}_{#1}^{#2}}

\sym{gfch}[O{}O{}]{\mathit{GFCN}_{#1}^{#2}}
\name{gfch}{generation-FACTS-choice network}{GFCN}

\name{dc}{Linear DC network}{LDC network}
\name{fdc}{FACTS Linear DC network}{FLDC network}

\sym{dsol}{(\pa, \flow, \gen, \load)}
\name{optsol}{optimal solution}{OS}

\newtheorem{theorem}{Theorem}[section]
\newtheorem{lemma}[theorem]{Lemma}
\newtheorem{definition}{Definition}

\newif\ifcomments \commentstrue
\newcommand\checkspace{\ifvmode\vspace{80pt}\penalty0\vspace{-80pt}\leavevmode\fi}
\newcommand\defedit[3]{\newcommand#1[1]{\ifcomments{\color{#2}{\checkspace
                                                    \ifinner{{\color{#2}{[#3]: }}}\else\marginpar{{\color{#2}{#3}}}\fi
                                                    \bf{##1}}}\fi}}
\defedit{\rbc}{red}{RB}
\defedit{\fpc}{green}{FP}
\defedit{\klc}{cyan}{KL}

\author{Karsten Lehmann, Russell Bent and Feng Pan}
\title{\LARGE \bf Maximizing electrical power supply using FACTS devices}

\begin{document}
\maketitle
\thispagestyle{empty}
\pagestyle{empty}

\begin{abstract}
Modern society critically depends on the services electric power provides.  Power systems rely on a network of power lines and transformers to deliver power from sources of power (generators) to the consumers (loads). However, when power lines fail (for example, through lightning or natural disasters) or when the system is heavily used, the network is often unable to fulfill all of the demand for power. While systems are vulnerable to these failures, increasingly, sophisticated control devices are being deployed to improve the efficiency of power systems. Such devices can also be used to improve the resiliency of power systems to failures.  In this paper, we focus on using FACTS devices in this context. A FACTS  device allows power grid operators to adjust the impedance parameters of power lines, thereby redistributing flow in the network and potentially increasing the amount of power that is supplied. Here we develop new approaches for determining the optimal parameter settings for FACTS devices in order to supply the maximal amount of power when networks are stressed, e.g. power line failures and heavy utilization.
\end{abstract}

\IEEEpeerreviewmaketitle

\section{Introduction}

One of the challenges facing modern infrastructures is the possibility of events that cause large-scale damage.  These events include hurricanes (i.e. Superstorm Sandy, Hurricane Katrina, etc.), ice storms, earthquakes, etc. During these events, significant portions of a society may be without services, such as electric power, that these systems provide.  As a result, there is considerable interest in developing approaches and methods for mitigating the effects of these events and restoring these systems as quickly as possible \cite{VanHentenryck2011,Adibi1994}.  One approach for mitigating the effects utilizes controls in the system to limit the impact of large-scale events. While installing such controls solely for the purpose of responding to big events may be too expensive in general, many systems have experienced a rapid rise in deployment of advanced control technologies to improve efficiency, cost, and reliability to small-scale events \cite{UnitedStatesDepartmentofEnergy2012}. Here, we focus on how existing adopted technologies may be used during emergency situations to mitigate the impacts of large-scale events.  In particular, we focus on how Flexible AC Transmission System (FACTS) \cite{hingorani2000understanding} improve power throughput when a system is stressed due to damage or increased utilization of the system.  

The flow of electric power in power systems is governed by complicated non-linear physics.  From a practical perspective, this means power flows along paths of least resistance from sources of power to consumers of power.  As a result, operators have limited ability to direct how power flows in a network.  This lack of control has a number of consequences, but the most important consequence here is that a network's capacity may be underutilized.  One solution to this problem is FACTS devices.  Using a wide variety of technologies, FACTS devices allow an operator to modify the resistivity of power lines, to shift power away or towards portions of a network. As a result, power systems may use more of their capacity, are more evenly utilized, and less expensive power generation may be dispatched.  While their primary use is in daily operations to improve efficiency, security, and economics, FACTS devices, as shown in this paper, are also useful for improving system response during stressed and adverse conditions.

In general, most recent work of FACTS device related optimization and control has focused on how FACTS can contribute to the stability \cite{oudalov2001application}, \cite{wu2003facts}, \cite{glanzmann2005using}, \cite{rehtanz2008coordinated}, \cite{sauvain2008optimal} and the maximum load-ability of a network \cite{orfanogianni2003steady}, \cite{feng2001allocation}, \cite{gotham1998power}, \cite{santiago2006optimal}. There is also work
 identifying where to place FACTS devices such that the load-ability of a network is maximized \cite{gerbex2001optimal}, \cite{lee2007hybrid}, \cite{wu2008optimal}.
\cite{jordehi2012approaches} provides a survey of different goals and methods that are used in FACTS optimization.
All of those approaches use heuristics (for example genetic algorithms, particle swarm) and run simulations on small power networks (up to 59 buses).
An exception to these papers is the work \cite{frolov2013reinforcing} that studies the FACTS placement problem on the 2736 bus polish network using the Linear DC model. 
In this paper we also use the Linear DC model and in contrast to existing work, we study the maximum throughput of a network on the big 2736 buses polish network. We also focus on developing a globally optimal approach as opposed to heuristics.  

The key contributions of this paper are as follows:
\begin{itemize}
\item An optimization model for maximizing power throughput in power systems when FACTS devices are available.

\item Empirical results demonstrating how FACTS devices increase power throughput in stressed and damaged systems.  

\item A discussion of the computational complexity of maximizing power throughput in power systems with FACTS devices.

\item Unlike similar problems, such as transmission switching \cite{Fisher2008},  our empirical results suggest that, in practice, the problem is often tractable to solve to small optimality gaps, even for large systems.  

\item An algorithm for improving the computational performance of commercial solvers on this problem.

\end{itemize}
The remainder of this paper is organized as follows.  Section II defines the problem. Section III discusses the complexity of the problem and Section IV discusses our algorithm for solving the problem. Section V describes the results of our computational experiments and we conclude with Section VI.


\section{Problem Definition}
\label{model}

Power flows in electrical power networks are described by nonlinear, steady-state electrical power flow equations (Alternating Current Model, AC).
In this paper we use the \emph{Linear DC (LDC) model} which is an approximation (linearization) of the AC model \cite{Schweppe_1970_Powersystemstatic}.
This model ignores reactive power and resistance and assumes that all voltages magnitudes are one in the per-unit system.
What remains are susceptances (the negative inverse of the reactance), the capacities and the phase angles of the voltages.
The flow on a line is similar to that of DC currents.  
The susceptance is the counterpart to the DC resistance, phase angles are the counterpart to the DC voltages and the power is the counterpart to the current.
FACTS devices are physical devices allowing the (otherwise constant) susceptance parameter to vary.

\begin{definition}
	A \emph{\fdcn (\fdca)} is a tuple $\net = (\buses, \generators, \loads, \lines)$ where \buses is the set of \emph{buses}; $\generators \subseteq \buses$ is the set of \emph{generators}; $\loads \subseteq \buses \setminus \generators$\footnote{W.l.o.g. buses with load and generation can be split into separate buses.} is the set of \emph{loads}; and $\lines \subseteq \powerset{\buses}[2] \times \intervalsp \times \preals$ 
is the set of \emph{lines} with their \emph{susceptance limits} and \emph{capacity}.\footnote{W.l.o.g. we assume that no two lines connect the same pair of buses.}
\end{definition}
A \emph{\dcn (\dca)} network is an \fdca without FACTS devices, i.e. all susceptances are fixed.
We define functions \func{\suscap}{\lines}{\intervalsp} for the susceptance limits and \func{\capa}{\lines}{\preals} for the capacities of power lines.
For a line from $a$ to $b$ with susceptance limit $[s,t]$ and capacity $p$ we use notation \edge[ab][[s,t]][p].
If the susceptance is fixed, i.e., $t = s$, we write \edge[ab][s][p]. We may also ignore these values and simply refer to the line by \edge[ab].
While this model does not explicitly give upper bounds on the generation or load of a bus,  
such constraints can be modeled by connecting these buses to the network 
through a single line whose capacity is the maximum output/intake of the bus.  

We now introduce the notations and equations describing  \fdca power flows.
We assume a fixed \fdca $\net = (\buses, \generators, \loads, \lines)$.
The \emph{generation} and \emph{load} at a bus are given by functions \func{\gen}{\buses}{\preals} and \func{\load}{\buses}{\preals} such that $\forall a \in \buses \setminus \generators: \gen[a] := 0$ and $\forall a \in \buses \setminus \loads: \load[a] := 0$. 
Also, we define functions \func{\sus}{\lines}{\preals} and \func{\pa}{\buses}{\reals} such that $\sus[e]$ is the \emph{susceptance} of line $e$ and $\pa[a]$ is the 
\emph{phase angle} at bus $a$.
The \emph{flow} on a line is given by function \func{\flow}{\lines}{\reals}. 
While the lines of \fdcas are undirected, orientation is needed to describe flows.
However, the concrete orientation we choose does not influence the theory.
To that end, whenever we define a line, we abuse the notation \edge[ab] to indicate that $\flow[ab] \geq 0$ whenever the flow goes from $a$ to $b$ and $\flow[ab] \leq 0$ otherwise.

The \dca model imposes two laws: Kirchhoff's conservation law and the \dca power law.

\begin{definition}
  A triple $(\flow, \gen, \load)$ satisfies \emph{Kirchhoff's conservation law} if $\forall a \in \buses:
  \sum_{\edge[ab] \in \lines} \flow[ab] - \sum_{\edge[ba] \in \lines} \flow[ba] = \gen[a] - \load[a].$
\end{definition}

\begin{definition}
    A triple $(\sus, \pa, \flow)$ satisfies the \emph{\dca power law} if $\forall \edge[ab] \in \lines: \flow[ab] = \sus[ab] (\pa[b] - \pa[a])$
    \footnote{Normally, the susceptance is a negative value and the flow equation is $\sus[ab](\pa[a] - \pa[b])$. For notation simplicity, we make the susceptance a positive value and multiply the flow equation by $-1$.}.
\end{definition}

\begin{definition}
    We call a tuple \sol a \emph{\soln} if: $(\flow, \gen, \load)$ satisfies Kirchhoff's conservation law; $(\sus, \pa, \flow)$ satisfies the \dca power law; $\forall e \in \lines: \sus[e] \in \suscap[e]$ and $\forall e \in \lines: |\flow[e]| \leq \capa[e]$.\\
Finally, we use \sols for the set of all \solns of \net.
\end{definition}

The \mfn \mf of a network \net is a triple $(\flow, \gen, \load)$ that maximizes the generation w.r.t. respecting Kirchhoff's conservation law; the line capacities and the generation and load bounds.
%
We now define two variants of this problem for power networks.
The \emph{\mffn (\mffa)} additionally has to satisfy the LDC power law. 
The \emph{\mpfn (\mpfa)} is a variant of \mffa where susceptance is fixed (i.e., it applies to an \dca).

\begin{definition}
The \emph{\mffn (\mffa)} of an \fdca \net is defined as
$\mff := \max_{\sol \in \sols} \sum_{g \in \buses} \gen[g].$

The \emph{\mpfn (\mpfa)} of an \dca \net is defined as
$\mpf := \max_{\dsol \in \sols} \sum_{g \in \buses} \gen[g].$
\end{definition}

Figure~\ref{triangle} shows an \dca where $g$ is a generator (box), $l$ is a load (house) and $b$ is a bus (sphere). 
We omit the susceptance and capacity of a line when its value is $1$.
Here, the \mfa for this network is $34$ whereas in the LDC model, we only can supply $16$ as shown in Figure~\ref{triangle_mpf} because the congestion of the edge \edge[bl][1][4] constrains the phase angle (written as $A=$ in the buses) between $g$ and $l$. 
Figure~\ref{triangle_facts} shows a variant of the network with two FACTS devices. These devices allow the maximum generation to reach $28$ as shown in Figure~\ref{triangle_mff}.
\pics[Examples for \mffa and \mpfa.]{
	\subpic{An \dca.}{triangle}
	\subpic{The \mpfa.}{triangle_mpf}
	\subpic{An \fdca network.}{triangle_facts}
	\subpic{The \mffa.}{triangle_mff}		
}[pic:examples]

\section{Computational Complexity}
Finding the solution to the \mpfa is known to be polynomial as it can be described as a linear program (LP).
In our workshop paper \cite{lehmann2014} we prove that finding the solution to the \mffa is NP-complete even for simple network structures.
For completeness, we discuss the main idea of the proof here.
We also present three special cases where the problem is polynomial: the network is a tree; all lines \edge[ab][[s,t]][p] have FACTS devices with either $s=0$ or $t=\infty$.
In all three cases the \mffa is equal to the \mfa.

\begin{lemma}
    Let \net be an \fdca with a tree structure then $\mff = \mf$.
\end{lemma}
\begin{proof}
    This is a consequence of the absence of cycles.
    Hence there are no cyclic dependencies on the phase angles which allows us to chose them in a way to match any optimal solution of the traditional max flow.
\end{proof}

\begin{lemma}
    \label{lem:szero}
    Let \net be an \fdca where for all lines \edge[ab][[s,t]][p]: $s = 0$ then $\mff = \mf$.
\end{lemma}
\begin{proof}
Let $(\flow, \gen, \load)$ be an acyclic \optsoln of the \mfa.
We have to define susceptance $\sus \in \suscap$ and phase angles \pa such that the DC power law is satisfied.
First, we define preliminary phase angles $\pa'$ using arbitrary positive values with the restriction that they respect the flow directions, so $\forall \edge[ab] \in \lines: 
    \pa[a]['] > \pa[b]['] \text{ if } \flow[ab] < 0; 
    \pa[a]['] < \pa[b]['] \text{ if } \flow[ab] > 0; \text{ and }
    \pa[a]['] = \pa[b]['] \text{ if } \flow[ab] = 0.$
    This defines susceptances $\sus'$ via the DC power law: $\forall \edge[ab] \in \lines: \sus[ab]['] := \frac{\flow[ab]}{\pa[b]['] - \pa[a][']}$. 
We now scale these susceptances such that they fit into their limits.
Let $x := min \{ \frac{s}{\sus[ab][']} \mid \edge[ab][[0,s]][p] \in \lines\}$.
By setting $\sus := x \sus'$ and $\pa := \frac{1}{x} \pa'$ we obtain $\forall \edge[ab] \in \lines: \flow[ab] =  \sus[ab]['] (\pa[b]['] - \pa[a][']) = x\sus[ab]['] (\frac{1}{x}\pa[b]['] - \frac{1}{x}\pa[a][']) = \sus[ab] (\pa[b] - \pa[a])$
and hence $(\sus, \pa, \flow)$ satisfies the LDC power law. 
Also, using the definition of $x$, for an arbitrary $\edge[ab][[0,s]][p] \in \lines$ we have $x \leq \frac{s}{\sus[ab][']}$ and hence $0 \leq \sus[ab] \leq s$. 
\end{proof}

\begin{lemma}
    Let \net be an \fdca where for all lines \edge[ab][[s,t]][p] we have $t = \infty$ then $\mff=\mf$.
\end{lemma}
\begin{proof}
    Similar to the proof of Lemma \ref{lem:szero}.
\end{proof}

In \cite{lehmann2014} we prove that deciding whether or not $\mff \geq x$ for some given $x \in \reals$ is strongly NP-complete in general and NP-complete when we restrict the network structure to cacti.
A network is a cactus if every line is part of at most one cycle.
The key element of these proofs are \emph{choice networks}. 
While the workshop paper presents the pure mathematical proof of the properties of choice networks, the reminder of this section presents the underlying idea that makes these properties possible.

Choice networks are used to encode (discrete) choices that characterize NP-hard problems.
Given an $x \in \preals$, 
a choice network can be regarded as a black box with port $p$ where we have external generation of either $x$ or $0$ at $p$ in order to achieve the inner maximum generation.
External generation indicates that the black box acts as generator for the network that is connected to to $p$.
Inner generation is the generation produced inside the black box.
\pic[A choice network with port $p$.]{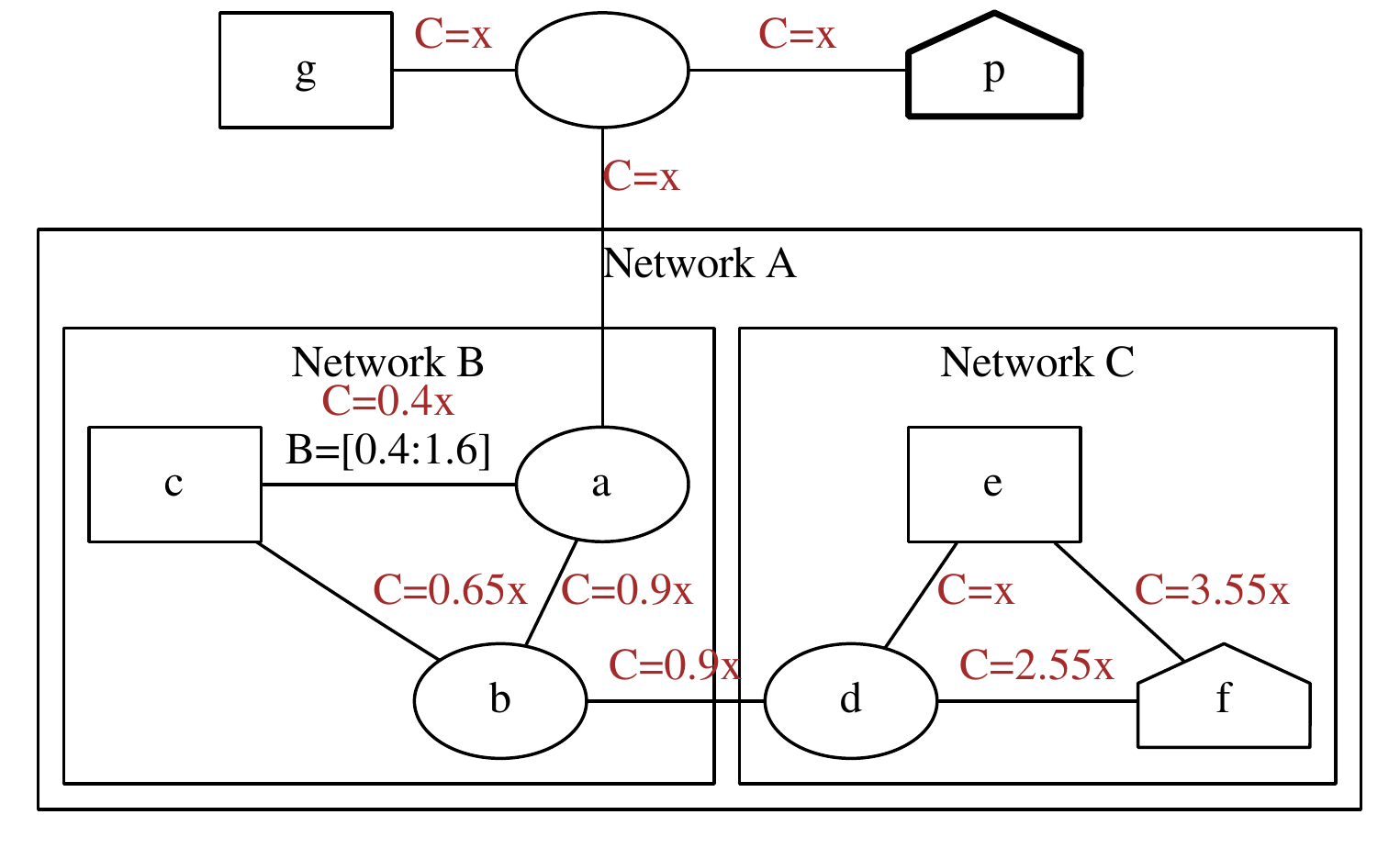}
Figure~\ref{gfch} presents our generation-FACTS-choice network \gfch[p,x].

In the following we describe the way the choice network works for the case $x=1$.
In Figure~\ref{gfch}, the generator can deliver 1 unit of  power.
Let the flow to $p$ be $w$ and the flow to the network $A$ be $r$, so that $w + r = 1$.
For the port $p$ we have $\mff[p] = w$.
The \mffa of port $p$ depends on $w$ according to a linear function of the form $f(z) = n +mz$.
Here, $n$ is the \emph{base generation} and $m$ is the \emph{generation ratio}.
The port $p$ has a base generation of $0$ and a generation ratio of $1$.
In the DC model and in a network without FACTS device, $n$ and $m$ are independent from the input.
The usage of FACTS devices allows us to construct networks where both values change if the input exceeds some threshold.
As we will show later, network $A$ is constructed in a way that $\mff[A] = 5.1 + \frac{2}{3}r$ if $r \leq 0.65$ and $\mff[A] = (5.1-\frac{1}{3}) + \frac{4}{3}r$ otherwise.
Using the equation $w = 1-r$ and assuming $r \leq 0.65$ we have $\mff[\gfch[p,1]] = 1 - r + 5.1 + \frac{2}{3}r = 6.1 - \frac{1}{3}r$ which has a single maximum of $6.1$ at $r=0$.
On the other hand, if $r \geq 0.65$ then we have $\mff[\gfch[p,1]] = 1 - r + (5.1-\frac{1}{3}) + \frac{4}{3}r = 6.1 - \frac{1}{3} + \frac{1}{3}r$ which has a single maximum of $6.1$ at $r=1$. 
This implies that there are exactly two solutions to achieve a flow of $6.1$ and that port $p$ has $0$ or $1$ units of power.
Looking at the general case, we see that $p$ either generates $0$ or $x$.

The \mffa of $A$ depends on the value of $r$: $\mff[A] = 5.1 + \frac{2}{3}r$ if $r \leq 0.65$ and $\mff[A] = (5.1-\frac{1}{3}) + \frac{4}{3}r$ otherwise.
The key feature to make it possible that there are exactly two solutions for the \mffa is that the generation ratio changes from a value less than one ($\frac{2}{3}$) to a value greater than one ($\frac{4}{3}$).
We now explain why $A$ has two generation ratios.
The network $A$ consists of two networks in sequence: $B$ and $C$.
The network $C$ has a fixed generation ratio of $2$. 
Every unit of power that enters $d$ has to go to $f$.
This increases the phase angle difference between $d$ and $f$ by one.
Hence the phase angle difference between $e$ and $f$ increases by one and gives us an additional unit of power.
Network $B$ has two generation ratios: $\frac{1}{3}$ for an input less then $0.65$ and $\frac{2}{3}$ otherwise.
Because $B$ and $C$ are in sequence, the ratios multiply and we get ratios of $\frac{2}{3}$ and $\frac{4}{3}$ for $A$.

Using choice networks, we reduced the exact cover by 3-set problem into the \mffa problem \cite{lehmann2014}.
Figure~\ref{theoencoding} presents the idea of the proof.

\pic[Example encoding for $(M,S) = (\{a,b,c,d,e,f\},\allowbreak \{\{a,b,c\}, \{b,c,d\},\allowbreak \{d,e,f\}\})$.]{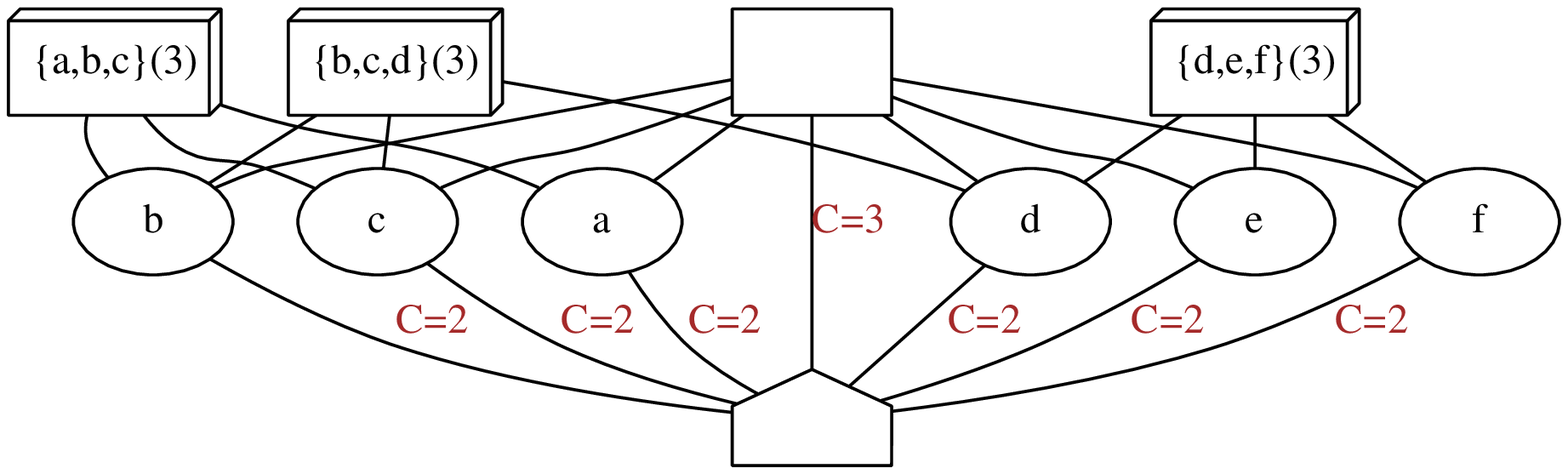}
\begin{theorem}
    \label{theoencoding}
    Let $x \in \reals$ and \net be an \fdca. Deciding if $\mff \geq
    x$ is strongly NP-complete.
\end{theorem}
\begin{proof}
    We prove this by reduction from the \emph{exact cover by 3-set problem}. Given a set $M$ and a set of subsets $S \subseteq \powerset{M}$ where every element of $S$ has exactly 3 elements, decide if there exist a set $T \subseteq S$ such that $\bigcup_{X \in T} X = M$ and $\forall X_1, X_2 \in T: X_1 \neq X_2 \implies X_1 \cap X_2 = \emptyset$.

For an instance $(M,S)$, we define the \fdca $\net[*][M,S] := (\buses, \generators, \loads, \lines)$ with $\generators := \{g\}$, $\loads := \{l\}$, $\buses := \generators \cup \loads \cup \bigcup_{X \in S} \{v_X\} \cup \bigcup_{x \in M} \{x\}$ and 
$\lines := \{\edge[gl][1][3]\} \cup \bigcup_{X \in S} \bigcup_{x \in X} \{\edge[v_Xx][1][1]\}) \cup \bigcup_{x \in M} \{ \edge[gx][1][1], \edge[xl][1][2]\}.$
We then define $\net[][M,S] = \net[*][M,S] + \sum_{X \in S} \gfch[3, v_X]$ and we have:
$\mff[\net[][M,S]] = 3 + 18.3|S| + |M| \iff (M,S) \text{ is solvable}.$
An example encoding for the exact cover problem $(M,S) = (\{a,b,c,d,e,f\},\allowbreak \{\{a,b,c\},\allowbreak \{b,c,d\},\allowbreak \{d,e,f\}\})$ can be found in Figure~\ref{exact_cover}.
The choice networks are represented by 3D boxes where the number $3$ in parenthesis is the value chosen for $x$.

For each $b \in M$ we have a bus $b$.
The network is constructed such that $\mff[\net[][M,S]] = 3 + 18.3|S| + |M|$ iff all choice networks have their inner maximum generation ($3\cdot6.1$) and all lines from the generator are congested.
This implies a phase angle difference of $3$ between the generator and the load and a phase angle difference of $1$ between $b$ and the generator.
Hence we have a phase angle difference of $2$ between the load and $b$.
For $b$ to satisfy Kirchhoff's conservation law, $b$ receives one unit of flow from a \gfch[3, v_X] with $b \in X$.
Because \gfch[3, v_X] can only generate $3$ if it wants to generate anything, all $c \in X$ get one.
Hence, we can only achieve the proposed \mffa value iff $(M,S)$ is solvable.
\end{proof}

\section{Global Optimization}

In this section we present a mixed integer program (MIP) model to find the solution to the \mffa.
The main problem in solving the \mffa is the right-hand-side of the DC power law $\flow[ab] =  \sus[ab] (\pa[b] - \pa[a])$.  This constraint contains the product of two variables which, in general, is a non-convex quadratic constraint (as implied in the previous section). 

However, for solving the \mffa, we do not need to know the susceptance values, we only need to ensure that there exists valid susceptance values such that the phase angle difference and the line flow are bound together. 
The solution region for just the phase angle difference and the flow consists of two convex regions.
There is one region for positive phase angle differences and one region for negative phase angle differences.
Given an edge \edge[ab][[s,t]][p] the positive region is described by $s(\pa[b] - \pa[a]) \leq \flow[ab] \leq t(\pa[b] - \pa[a])$.
To formulate the problem as a MIP, we introduce a binary variable $d_{\edge[ab]}$ that represents the choice of phase angle difference direction.
Assuming that $M$ is some upper bound for the phase angle difference, the MIP model can be found in Model \ref{mip}.
\begin{Model}
    \caption{MIP model to solve the \mffa.}
        \label{mip}
\begin{align}
& {\text{maximize}}
& & \sum_{a \in \buses} \gen[a] \label{eqn:obj}\\
& \text{subject to}
& & d_{\edge[ab]} \in \{0,1\}, \; \edge[ab] \in \lines \label{constraint:discrete} \\ 
&&& 0 \leq \padiff[ab][+] \leq d_{\edge[ab]} M, \; \edge[ab] \in \lines \label{constraint:pos}\\ 
&&& 0 \leq \padiff[ab][-] \leq (1-d_{\edge[ab]}) M, \; \edge[ab] \in \lines \label{constraint:neg}\\ 
&&& \padiff[ab][+] - \padiff[ab][-] = \pa[b] - \pa[a], \; \edge[ab] \in \lines \label{constraint:phasediff}\\ 
&&& s \padiff[ab][+] \leq \flow[ab][+] \leq t \padiff[ab][+], \; \edge[ab][[s,t]][p] \in \lines \label{constraint:posflow}\\
&&& s \padiff[ab][-] \leq \flow[ab][-] \leq t \padiff[ab][-], \; \edge[ab][[s,t]][p] \in \lines \label{constraint:negflow}\\
&&& \flow[ab] = \flow[ab][+] - \flow[ab][-], \; \edge[ab] \in \lines \label{constraint:flow}\\ 
&&& \sum_{\edge[ab] \in \lines} \flow[ab] - \sum_{\edge[ba] \in \lines} \flow[ba] = \gen[a] - \load[a], \; a \in \buses \label{constraint:balance}\\
&&& -\capa[ab] \leq \flow[ab] \le \capa[ab], \; \edge[ab] \in \lines \label{constraint:cap}
\end{align}
\end{Model}

\noindent
In this model, Equation \ref{eqn:obj} describes the objective function that maximize the amount of flow in the system by maximizing the total generation. Equation \ref{constraint:discrete} constrains the flow direction variables as binary. Then, equations \ref{constraint:pos} and \ref{constraint:neg} force either the positive or negative phase angle difference of $ab$ to be 0 depending on the choice of $d$. Equation \ref{constraint:phasediff} states that the difference in phase angles (right hand side), as stated by $\Theta$, is equal to the sum of the negative and positive phase angle differences as stated by $\Delta$.  This constraint couples the flow direction (edge based variables) with the phase angles (node based variables). Equations \ref{constraint:posflow} and \ref{constraint:negflow} state that the flow in the positive or negative direction can be any value that is consistent with the phase angle differences and the bounds on the susceptances. Equation \ref{constraint:flow} states the flow on $ab$ is the sum of the negative and positive flows. Equation \ref{constraint:balance} states that flow must be balanced at each node. Finally, equation \ref{constraint:cap} constrains the flow on $ab$ to be smaller than its capacity.  After this problem is solved, the susceptance of $ab$ is calculated as $\sus[ab] = \frac{\flow[ab]}{\pa[b] - \pa[a]}$.

During our experimental testing, we observed cases where commerical solvers such as Gurobi and Cplex were unable to find provably good solutions on large problems. After one CPU hour the optimality gap was larger than 400\%, in some cases.  Based on these observations, we developed an approach to boost the performance of the solvers by providing high quality initial solutions.  We refer to this approach as the  \emph{iterative method (IM)}.
The IM is based on two observations. 
First, finding the \mpfa is polynomial.
That means that finding the maximum flow in a network with fixed susceptances is easy.
Second, if we fix the binary variables in the MIP model, we are left with an LP.
Fixing the binary variables is equivalent to fixing the phase angle difference for all lines.
We refer to the maximum possible flow possible when the sign of the phase angle differences fixed  as the \mvfn (\mvfa).
The IM works as follows, given some random valid susceptances (\sus) we find the \mpfa.
Then we take the sign of the phase angle differences ($D$) from this solution, fix them and solve the \mvfa problem.
From this solution we take the susceptances (\sus) and put them into the \mpfa model to start the next iteration.
This process is described in Algorithm \ref{iterative}.
We continue iterating until we converge to a solution.
\begin{algorithm}
    \KwData{susceptances \sus}
    \KwResult{phase angle difference directions $D$}
    MPF = -1\;
    MVF = 0\;
    \While{MPF $<$ MVF}{
        MPF, $D$ = $solveMPF(\sus$)\;
        MVF, \sus = $solveMVF(D)$\;
    }
    \caption{The iterative method (IM).}
    \label{iterative}
\end{algorithm}

\section{Experimental Results}

For our experimental results we focused on evaluating the computation required for solving the \mpfa, both with and without the IM.  We also compared the solutions obtained by the \mpfa with the solutions of \mffa in order to provide some evidence of the types of solution improvement one might expect when FACTS devices are used.  We used the IEEE test problems provided with Matpower \cite{Zimmerman2011} to test our approach.  In general, on smaller networks, the \mffa is computationally easy to solve (under 1 CPU minutes) and the differences between \mpfa and \mffa are small.  Thus, we focus on presenting results produced on the IEEE 2736 bus system (based on Poland's power grid) due to its large size and computational challenges.  The results were obtained on a computer with  an Intel Core i7-4702HQ Quad Core processor using Gurobi 5.5 \cite{gurobi}.

We used two approaches to create variants of the IEEE problems that mimic damage and heavy utilization.
Our first method removes random lines to simulate line failures.
\pic[IEEE 2736 damage scenarios that remove randomly selected lines (x axis).  The y-axis shows the maximum throughput for  \mpfa and  \mffa.]{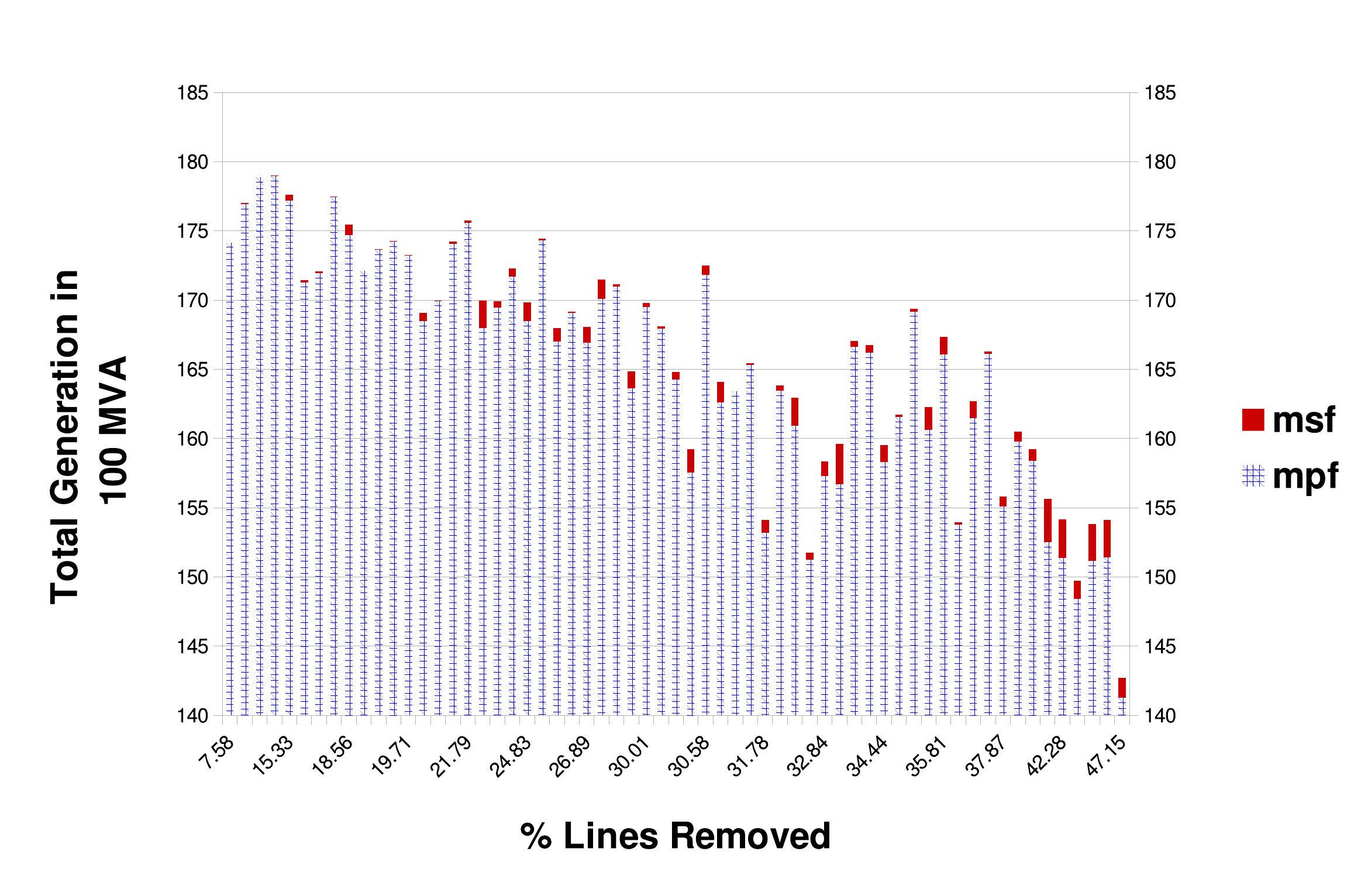}[scale=0.35]
Figure~\ref{line_removal} shows results on  IEEE 2736 for 60 scenarios where random sets of lines are removed from the power network.  The graphs compare the quality of the solution for \mpfa with  \mffa .
 \mffa was initialized with IM and was allowed to run for 10 CPU minutes.
In each case we selected a random set of lines that have FACTS devices according to a uniform distribution. Each device was allowed to vary the suceptances $30\%$.  With these results, we see that as the amount of damage increases, the benefits provided  by \mffa increase (up to  $5\%$).

\begin{table*}
\caption{A comparison of MFF and MPF on IEEE2736 for different congestion factors.  $30\%$ of the lines have FACTS devices}

\center
\label{tab2736_cf}
\begin{tabular}{|c|c|c|c|c|c|c|c|c|c|c|c|c|}
\hline
Generation & \multicolumn{3}{|c|}{1.5}&\multicolumn{3}{|c|}{2.0}&\multicolumn{3}{|c|}{2.5}&\multicolumn{3}{|c|}{3.0}\\
\hline
Load & \mpfa & \mffa & Improv.& \mpfa & \mffa & Improv.& \mpfa & \mffa & Improv.& \mpfa & \mffa & Improv.\\
\hline\hline
1.5&270.56&271.0&0.16&271.12&271.12&0.0&271.12&271.12&0.0&271.12&271.12&0.0\\
\hline
2.0&303.05&303.13&0.03&347.8&353.47&1.63&354.99&358.68&1.04&358.17&360.33&0.6\\
\hline
3.0&303.13&303.13&0.0&392.83&401.49&2.2&433.85&445.88&2.77&450.15&462.18&2.67\\
\hline
4.0&303.13&303.13&0.0&401.26&403.12&0.46&461.17&473.41&2.66&490.41&502.01&2.36\\
\hline
\end{tabular}
\end{table*}

\begin{table*}
\caption{Results comparing \mffa with IM on problem IEEE2736 with different FACTS devices. The \mpfa of the network is $419.19$. IM significantly improves the quality of solutions.}
\label{tab2736}
\center
\begin{tabular}{|c|c|c|c|c|c|c|c|c|c|c|c|}
\hline
\multicolumn{2}{|c|}{Net} & \multicolumn{3}{|c|}{\mffa (1h time limit)} & \multicolumn{3}{|c|}{3-IM} & \multicolumn{3}{|c|}{3-IM + \mffa (10min time limit)} & Improv.\\
\hline
FACTS\% & Interval\% & Obj & RT & MIPgap & Obj & RT & Calls & Obj & RT & MIPgap & to \mpfa in \%\\
\hline\hline
30 & 40 & 405.32 & 3600 & 15.69
& 463.83 & 306 & 83.00
& 463.86 & 600 & 1.29
& 10.66
\\
\hline
60&70& 455.22 & 3600 & 3.07
& 466.87 & 320 & 91.00
& 466.98 & 600 & 0.71
& 11.4
\\
\hline
60&10& 91.47 & 3600 & 398.14
& 445.78 & 191 & 45.00
& 445.80 & 600 & 2.34
& 6.35
\\
\hline
10&40& 449.72 & 3602 & 4.25
& 458.21 & 188 & 23.00
& 460.21 & 600 & 2.13
& 9.79
\\
\hline
30&10& 88.85 & 3600 & 424.37
& 457.66 & 224 & 41.00
& 457.69 & 600 & 1.80
& 9.18
\\
\hline
60&40& 343.99 & 3603 & 35.85
& 460.12 & 218 & 41.00
& 460.15 & 600 & 1.62
& 9.77
\\
\hline
90&40& 376.41 & 3600 & 23.58
& 451.43 & 176 & 21.00
& 454.08 & 600 & 2.52
& 8.32
\\
\hline
90&70& 455.22 & 3600 & 3.07
& 466.87 & 320 & 91.00
& 466.98 & 600 & 0.71
& 11.4
\\
\hline
10&10& 385.63 & 3600 & 21.42
& 463.98 & 304 & 73.00
& 464.04 & 600 & 1.07
& 10.7
\\
\hline
10&70& 455.22 & 3600 & 3.07
& 466.65 & 230 & 67.00
& 466.73 & 600 & 0.55
& 11.34
\\
\hline
30&70& 455.22 & 3600 & 3.07
& 466.87 & 254 & 65.00
& 466.98 & 600 & 0.71
& 11.4
\\
\hline
\end{tabular}
\end{table*}

Our second method scales the upper bound of load and generation to model increased utilization.
We refer to the scaling factors as generation and load \emph{congestion factors}, respectively.
Table~\ref{tab2736_cf} presents the solutions for \mpfa, \mffa and their improvement percentage for different congestion factors when $30\%$ of the lines (randomly chosen) have FACTS devices that
are allowed to vary from their initial susceptance by $30\%$.
\mffa was initialized with IM and was allowed to run for 10 CPU minutes.
The first row and column of this table note the generation and load congestion factors.  In this table, the results indicate that as the congestion factors increase, the benefit of \mpfa over \mffa increases by up to $3\%$.  

In most of the results discussed in this section, Gurobi is able to find solutions with a gap of $0.1\%$ in under an hour of CPU time without the need for IM. As part of our study, we generated a large number of random cases for different levels of FACTS penetration and different capabilities of FACTS to vary susceptances. In some of these cases, even with 1 CPU hour available,  Gurobi was unable to find reasonable solutions.  In general, we observed this behavior when the gap between \mpfa and \mffa was large (between 6 and 11\%).  We show some results for one case when the generation congestion factor was $2.375$  and the load congestion factor was $2.75$ in \ref{tab2736}.
The \mpfa of this network is $419.9$.
The first two columns of this table show the percentage of lines with FACTS devices and the allowed interval of susceptance (as a percentage of the original suspectance).
The \mffa column shows the results of \mffa without IM. 
The columns labeled Obj, RT, and MIPGap present the solution, run time in CPU seconds, and optimality gap, respectively.
The column 3-IM presents the results of the iterative method in terms of solution quality (IM), run time in CPU seconds (RT), and iterations (Calls).  Here, IM is executed 3 times with different starting solutions (susceptance upper bound, sucsceptance lower bound, and suscepatance mid point), and the best result is reported.
As seen in these results, the IM method produces a high quality solution in a short amount of time.
 In the 3-IM + MFF column, we show the results after 10 CPU minutes when IM is used as an initial solution for MFF.  Though the MIP solver is unable to improve the solutions much, it is able to show that the IM method (in most casess) produces solutions with substantially small optimality gaps and that are significantly better than \mpfa.

\section{Conclusion}
In this paper we discussed an optimization model for maximizing the throughput of a network utilizing FACTS devices and an algorithm that improves the computational performance of commercial solvers on this problem. In general, even for large scale problems, our approach is able to solve the problem to near optimality in under 10 CPU minutes.  On large scale problems, using FACTS devices when the network is over utilized or damaged is able to improve the throughput between 2 and $5\%$ (on occasion, higher).  While this appears to be a modest improvement, when discussing damage scenarios that result in millions of people without power (such as Hurricane Sandy), a $5\%$ improvement corresponds to a significant number of people.
We also provided a discussion that illustrates that the problem is NP-complete in general.

There remain a number of interesting future directions in FACTS device optimization for maximum throughput. It will be interesting to include other types of controls, such as switching, to determine if the combined control further improves system response.  It will also be interesting to consider problems that are of larger scale than the IEEE Polish systems.  Finally, future work should also conduct an in-depth study of problem structure that results in solutions with large gaps between MPF and MFF in order to guide the placement of FACTS devices.

\section*{Acknowledgment}
This work was partially supported by  the Los Alamos National Laboratory LDRD project \textit{Optimization and Control Theory for Smart Grids} and the Department of Energy Advanced Grid Science Program.
\bibliography{pes,ms,stateest,braess_paradox}

\begin{thebibliography}{10}
\providecommand{\url}[1]{#1}
\csname url@samestyle\endcsname
\providecommand{\newblock}{\relax}
\providecommand{\bibinfo}[2]{#2}
\providecommand{\BIBentrySTDinterwordspacing}{\spaceskip=0pt\relax}
\providecommand{\BIBentryALTinterwordstretchfactor}{4}
\providecommand{\BIBentryALTinterwordspacing}{\spaceskip=\fontdimen2\font plus
\BIBentryALTinterwordstretchfactor\fontdimen3\font minus
  \fontdimen4\font\relax}
\providecommand{\BIBforeignlanguage}[2]{{%
\expandafter\ifx\csname l@#1\endcsname\relax
\typeout{** WARNING: IEEEtran.bst: No hyphenation pattern has been}%
\typeout{** loaded for the language `#1'. Using the pattern for}%
\typeout{** the default language instead.}%
\else
\language=\csname l@#1\endcsname
\fi
#2}}
\providecommand{\BIBdecl}{\relax}
\BIBdecl

\bibitem{VanHentenryck2011}
P.~{Van Hentenryck}, C.~Coffrin, and R.~Bent, ``{Vehicle Routing for the Last
  Mile of Power System Restoration},'' in \emph{17th Power Systems Computation
  Conference}, Stockholm, Sweden, 2011.

\bibitem{Adibi1994}
M.~M. Adibi and L.~H. Fink, ``{Power System Restoration Planning},'' \emph{IEEE
  Transactions on Power Systems}, vol.~9, no.~1, pp. 22--28, 1994.

\bibitem{UnitedStatesDepartmentofEnergy2012}
{United States Department of Energy}, ``{2010 Smart Grid System Report},''
  Washington, DC, Tech. Rep., 2012.

\bibitem{hingorani2000understanding}
N.~G. Hingorani, L.~Gyugyi, and M.~El-Hawary, \emph{Understanding FACTS:
  concepts and technology of flexible AC transmission systems}.\hskip 1em plus
  0.5em minus 0.4em\relax IEEE press New York, 2000, vol.~1.

\bibitem{oudalov2001application}
A.~Oudalov, R.~Cherkaoui, and A.~Germond, ``Application of fuzzy logic
  techniques for the coordinated power flow control by multiple series facts
  devices,'' in \emph{Power Industry Computer Applications, 2001. PICA 2001.
  Innovative Computing for Power - Electric Energy Meets the Market. 22nd IEEE
  Power Engineering Society International Conference on}, 2001, pp. 74--80.

\bibitem{wu2003facts}
W.~Wu and C.~Wong, ``Facts applications in preventing loop flows in
  interconnected systems,'' in \emph{Power Engineering Society General Meeting,
  2003, IEEE}, vol.~1, July 2003, pp. --174 Vol. 1.

\bibitem{glanzmann2005using}
G.~Glanzmann and G.~Andersson, ``Using facts devices to resolve congestions in
  transmission grids,'' in \emph{CIGRE/IEEE PES, 2005. International
  Symposium}, Oct 2005, pp. 347--354.

\bibitem{rehtanz2008coordinated}
C.~Rehtanz and U.~Hager, ``Coordinated wide area control of facts for
  congestion management,'' in \emph{Electric Utility Deregulation and
  Restructuring and Power Technologies, 2008. DRPT 2008. Third International
  Conference on}, April 2008, pp. 130--135.

\bibitem{sauvain2008optimal}
H.~Sauvain, M.~Lalou, Z.~Styczynski, and P.~Komarnicki, ``Optimal and secure
  transmission of stochastic load controlled by wacs swiss case,'' in
  \emph{Power and Energy Society General Meeting - Conversion and Delivery of
  Electrical Energy in the 21st Century, 2008 IEEE}, July 2008, pp. 1--5.

\bibitem{orfanogianni2003steady}
T.~Orfanogianni and R.~Bacher, ``Steady-state optimization in power systems
  with series facts devices,'' \emph{Power Systems, IEEE Transactions on},
  vol.~18, no.~1, pp. 19--26, Feb 2003.

\bibitem{feng2001allocation}
W.~Feng and G.~Shrestha, ``Allocation of tcsc devices to optimize total
  transmission capacity in a competitive power market,'' in \emph{Power
  Engineering Society Winter Meeting, 2001. IEEE}, vol.~2, 2001, pp. 587--593
  vol.2.

\bibitem{gotham1998power}
D.~Gotham and G.~Heydt, ``Power flow control and power flow studies for systems
  with facts devices,'' \emph{Power Systems, IEEE Transactions on}, vol.~13,
  no.~1, pp. 60--65, Feb 1998.

\bibitem{santiago2006optimal}
M.~Santiago-Luna and J.~Cedeno-Maldonado, ``Optimal placement of facts
  controllers in power systems via evolution strategies,'' in
  \emph{Transmission Distribution Conference and Exposition: Latin America,
  2006. TDC '06. IEEE/PES}, Aug 2006, pp. 1--6.

\bibitem{gerbex2001optimal}
S.~Gerbex, R.~Cherkaoui, and A.~Germond, ``Optimal location of multi-type facts
  devices in a power system by means of genetic algorithms,'' \emph{Power
  Systems, IEEE Transactions on}, vol.~16, no.~3, pp. 537--544, Aug 2001.

\bibitem{lee2007hybrid}
K.~Lee, M.~Farsangi, and H.~Nezamabadi-pour, ``Hybrid of analytical and
  heuristic techniques for facts devices in transmission systems,'' in
  \emph{Power Engineering Society General Meeting, 2007. IEEE}, June 2007, pp.
  1--8.

\bibitem{wu2008optimal}
Q.~Wu, Z.~Lu, M.~Li, and T.~Y. Ji, ``Optimal placement of facts devices by a
  group search optimizer with multiple producer,'' in \emph{Evolutionary
  Computation, 2008. CEC 2008. (IEEE World Congress on Computational
  Intelligence). IEEE Congress on}, June 2008, pp. 1033--1039.

\bibitem{jordehi2012approaches}
A.~Jordehi and J.~Jasni, ``Approaches for facts optimization problem in power
  systems,'' in \emph{Power Engineering and Optimization Conference (PEDCO)
  Melaka, Malaysia, 2012 Ieee International}, June 2012, pp. 355--360.

\bibitem{frolov2013reinforcing}
V.~Frolov, S.~Backhaus, and M.~Chertkov, ``Reinforcing power grid transmission
  with facts devices,'' \emph{arXiv:1307.1940, preprint}, 2013.

\bibitem{Fisher2008}
E.~Fisher, R.~O'Neill, and M.~Ferris, ``{Optimal Transmission Switching},''
  \emph{IEEE Transactions on Power Systems}, vol.~23, no.~3, pp. 1346--1355,
  2008.

\bibitem{Schweppe_1970_Powersystemstatic}
F.~Schweppe and D.~Rom, ``Power system static-state estimation, part ii:
  Approximate model,'' \emph{power apparatus and systems, {IEEE} transactions
  on}, no.~1, pp. 125--130, 1970.

\bibitem{lehmann2014}
K.~Lehmann, A.~Grastien, and P.~Van~Hentenryck, ``The complexity of switching
  and facts maximum-potential-flow problems,'' in \emph{40th International
  Workshop on Graph-Theoretic Concepts in Computer Science}, under review.

\bibitem{Zimmerman2011}
\BIBentryALTinterwordspacing
R.~D. Zimmerman, C.~E. Murillo-Sanchez, and R.~J. Thomas,
  ``\BIBforeignlanguage{English}{{MATPOWER: Steady-State Operations, Planning,
  and Analysis Tools for Power Systems Research and Education}},''
  \emph{\BIBforeignlanguage{English}{IEEE Transactions on Power Systems}},
  vol.~26, no.~1, pp. 12--19, 2011. [Online]. Available: \url{<Go to
  ISI>://000286516100002}
\BIBentrySTDinterwordspacing

\bibitem{gurobi}
\BIBentryALTinterwordspacing
I.~Gurobi~Optimization, ``Gurobi optimizer reference manual,'' 2014. [Online].
  Available: \url{http://www.gurobi.com}
\BIBentrySTDinterwordspacing

\end{thebibliography}
\bibliographystyle{IEEEtran}

\begin{biographynophoto}{Karsten Lehmann}
    received Master degrees in Computer Science and Mathematics from Technische Universit\"at Dresden and is currently a PhD student at NICTA in the Optimisation Research Group and the Australian National University. He did a 3 month internship at LANL.\\[0.08in]

\setlength{\parindent}{0in}

{\bf Russell Bent}
received his PhD in Computer Science from Brown University in 2005 and is currently a staff scientist at LANL in the Energy and Infrastructure Analysis Group. His publications include deterministic optimization, optimization under uncertainty, infrastructure modeling and simulation, constraint programming, algorithms, and simulation. Russell has published 1 book and over 40 articles in peer-reviewed journals and conferences in artificial intelligence and operations research. \\[0.08in]

{\bf Feng Pan}
received his Ph.D. degree in Operations Research from the University of Texas, Austin, in 2005.
His research areas include combinatorial optimization, stochastic programming, and network modeling. 
He is a Research Scientist at Los Alamos National Laboratory, Los Alamos, NM. 
He has worked on various projects in transportation security and infrastructure modeling. 
He is currently working on optimization and control models in smart grids and is leading several basic research projects in the areas of developing network interdiction models and designing robust and resilient infrastructure networks.
 \\[0.08in]
\end{biographynophoto}
\end{document}